\documentclass[10pt]{article}

\usepackage{cmbright}
\usepackage[latin2]{inputenc}
\DeclareSymbolFont{operators}   {OT1}{cmr} {m}{n}
\DeclareSymbolFont{letters}     {OML}{cmm} {m}{it}
\DeclareSymbolFont{symbols}     {OMS}{cmsy}{m}{n}
\DeclareSymbolFont{largesymbols}{OMX}{cmex}{m}{n}
\SetSymbolFont{operators}{bold}{OT1}{cmr} {bx}{n}
\SetSymbolFont{letters}  {bold}{OML}{cmm} {b}{it}
\SetSymbolFont{symbols}  {bold}{OMS}{cmsy}{b}{n}
\DeclareSymbolFontAlphabet{\mathrm}    {operators}
\DeclareSymbolFontAlphabet{\mathnormal}{letters}
\DeclareSymbolFontAlphabet{\mathcal}   {symbols}
\DeclareMathAlphabet      {\mathbf}{OT1}{cmr}{bx}{n}
\DeclareMathAlphabet      {\mathsf}{OT1}{cmss}{m}{n}
\DeclareMathAlphabet      {\mathit}{OT1}{cmr}{m}{it}
\DeclareMathAlphabet      {\mathtt}{OT1}{cmtt}{m}{n}
\SetMathAlphabet\mathsf{bold}{OT1}{cmss}{bx}{n}
\SetMathAlphabet\mathit{bold}{OT1}{cmr}{bx}{it}

\usepackage{epsfig}
\usepackage{amsmath}
\usepackage{graphicx,psfrag}
\usepackage{dcolumn}
\usepackage{bm}
\usepackage{color}
\usepackage{slashed}

\newcommand{\hf} {\frac{1}{2}}

\newcommand{\nn}{\nonumber\\}

\newcommand{\dd}{\mathrm{d}}

\def\Tr{{\rm Tr}}

\def\eq#1{(\ref{#1})}
\def\s0#1#2{\mbox{\small{$ \frac{#1}{#2} $}}}
\def\0#1#2{\frac{#1}{#2}}

\def\mr#1{{\mathrm{#1}}}

\begin{document}

{\bf \large The Numerically Optimized Regulator and the Functional Renormalization Group}

{\sc I. G. M\'ari\'an$^{a}$, U. D. Jentschura$^{b,c}$ and I. N\'andori$^{c,d}$}

{\em a) University of Debrecen, P.O. Box 105, H-4010 Debrecen, Hungary}\\
{\em b) Missouri University of Science and Technology, Rolla, Missouri 65409-0640, USA}\\
{\em c) MTA-DE Particle Physics Research Group, P.O. Box 51, H-4001 Debrecen, Hungary}\\
{\em d) MTA Atomki, P.O. Box 51, H-4001 Debrecen,  Hungary}

{\small
We aim to optimize the functional form of the compactly supported smooth (CSS)
regulator within the functional renormalization group (RG), in the framework of
bosonized two-dimensional Quantum Electrodynamics (QED$_2$) and of the
three-dimensional $O(N=1)$ scalar field theory in the local potential
approximation (LPA). The principle of minimal sensitivity (PMS) is used for the
optimization of the CSS regulator, recovering all the major types of regulators
in appropriate limits. Within the investigated class of functional forms, a
thorough investigation of the CSS regulator, optimized with two different
normalizations within the PMS method, confirms that the functional form of a
regulator first proposed by Litim is optimal within the LPA. However, Litim's
exact form leads to a kink in the regulator function.  A form of the CSS regulator, 
numerically close to Litim's limit while maintaining infinite differentiability, remains 
compatible with the gradient expansion to all orders. A smooth analytic behaviour 
of the regulator is ensured by a small, but finite value of the exponential fall-off 
parameter in the CSS regulator. Consequently, a compactly supported regulator, 
in a parameter regime close to Litim's optimized form, but regularized with an 
exponential factor, appears to have favorable properties and could be used to 
address the scheme dependence of the functional renormalization group, at least 
within the the approximations employed in the studies reported here.}

\tableofcontents

\section{Introduction}
\label{sec_intro}
In particle physics, theories and models are defined at high energies, where 
symmetry considerations are valid, but the measurements are performed at 
relatively low energies. The low-energy behaviour of 
field-theoretical models is determined 
by the renormalization group. Usually, perturbative 
renormalization is sufficient. However, there are special situations, for example 
the confinement of quarks into hadrons, where non-perturbative treatments are 
needed. Renormalization can be performed non-perturbatively by means of the 
functional renormalization group (RG) method~\cite{WP,We1993,Mo1994,internal} 
which was applied successfully in many cases. The 
following functional RG equation for 
scalar fields~\cite{We1993} (Wetterich equation)
\begin{equation}
\label{erg}
k \partial_k \Gamma_k [\varphi] = \hf  \Tr \left[
(k\partial_k R_k) / (\Gamma_k^{(2)}[\varphi] + R_k)
\right]
\end{equation}
may be derived for the blocked effective action $\Gamma_k$ which interpolates 
between the bare action $\Gamma_{k\to \Lambda} \equiv S$ and the full quantum 
effective action $\Gamma_{k\to 0} \equiv \Gamma$ where $k$ is the running 
momentum scale. The second functional derivative of the blocked action 
is represented by $\Gamma_k^{(2)}$, and the trace Tr stands for the 
momentum integration. $R_k$ is the regulator function 
which is assumed to fulfill the conditions $R_k(p\to 0)>0$, 
as well as $R_{k\to 0}(p)=0$ and $R_{k\to \Lambda}( p)=\infty$. 
The physical results obtained by the exact RG equation are independent of the
particular choice of the regulator~\cite{litim_pawlowski}.

The RG equation constitutes a functional partial differential equation and it
is therefore not possible to indicate general solutions. Hence, approximations
are required. One of the commonly used systematic approximations is the
truncated gradient expansion \cite{scheme}.  The necessity of approximations
implies that the RG flow depends on the choice of the regulator function, i.e.,
on the renormalization scheme (``scheme dependence''). In principle, therefore,
physical results could become scheme-dependent. The comparison of results
obtained using various types of regulator functions (see
Refs.~\cite{opt_rg,litim_o(n),opt_func,Ro2010,Mo2005,qed2,scheme,scheme_sg,%
minimal_sens,css,opt_d1_d2,qeg_on_sanyi,d1_anharmonic}) is indicated. In order
to increase the predicting power of the RG method, an optimization of the
scheme-dependence is required. A rather general optimization procedure
(Litim--Pawlowski method~\cite{opt_rg,opt_func}) leads to Litim's
regulator~\cite{opt_rg}. In typical cases, it agrees well with experimental
data~\cite{litim_o(n)}, and furthermore, the corresponding RG equation can be
mapped onto the Polchinski RG at least in the leading order of the gradient
expansion \cite{Mo2005}. Its disadvantage is that it is non-differentiable and
thus incompatible with the gradient expansion~\cite{Mo2005,Ro2010}. Another
optimization scenario is based on the principle of minimal sensitivity (PMS,
see Ref.~\cite{minimal_sens}), where the optimal parameters of a given
regulator are chosen such as to make the physical quantities as insensitive as
possible to any conceivable changes of the parameters entering the regulator.
Its advantage is that it can be used at any order of the gradient expansion,
its disadvantage is that regulators of different functional form cannot easily
be compared to each other based on the PMS alone. The
``optimization'' then proceeds under the implicit assumption that the PMS
method converges to the optimized values rather than a non-optimal ``saddle
point'' in the parameter space. In the following, we reproduce a number of
known results for optimum parameters in limiting cases, and we reassure
ourselves that in all cases considered, the results of the PMS method are
consistent with the globally optimal values of the parameters that define the
regulator within the functional forms discussed in the current paper.

Recently, a new type of regulator function, the compactly supported smooth
(CSS) regulator~\cite{css}, has been introduced which encompasses all major
types of regulator functions discussed so far in the literature, in appropriate
limits. Thus, it can be used to compare various regulator functions to each
other in the framework of the PMS optimization method. Moreover, it is a
smooth, infinitely differentiable function, and it has a compact support (it is
non-zero only in a finite range). Therefore, it can be applied to consider the
``Litim limit'' at any order of the gradient expansion.

The purpose of this paper is to study the scheme-dependence of the functional
RG in the framework of the bosonized QED$_2$ (Ref.~\cite{qed2}) and the
$O(N=1)$ symmetric scalar field theory~\cite{litim_o(n)}, within the local
potential approximation (which is the leading order of the gradient expansion).
We use the CSS regulator function~\cite{css}.  Our goal is {\em (i)} to
optimize the parameters of the CSS regulator in LPA using the PMS optimization
procedure, and {\em (ii)} to compare the results obtained by the PMS strategy
to the one advocated by Litim.  For this, the CSS regulator has been
investigated with two types of normalization.

The organization of this paper is as follows.  In Sec.~\ref{sec_model}, wegive
a brief overview of the regulators and field-theoretical models studied here.
In particular, we consider typical approximations used in solving RG equations
and major types of regulator functions.  Furthermore, known results on QED$_2$
are summarized. In Secs.~\ref{sec_results_qed2} and~\ref{sec_phi4}, the
optimization of the CSS regulator is presented in the framework of QED$_2$ and
the $O(N=1)$ model, respectively. Conclusions are reserved for
Sec.~\ref{sec_sum}.

\section{Regulators and Models}
\label{sec_model}

\subsection{Approximations}
\label{sec_approx}
In order to solve the RG equation~\eq{erg},
one of the commonly used systematic 
approximations is the truncated gradient (i.e., derivative) expansion where 
$\Gamma_k$ is expanded in powers of the derivative of the field,
\begin{equation}
\label{deriv}
\Gamma_k [\varphi] = \int \dd^d x  \left[V_k(\varphi) 
+ Z_k(\varphi) \hf (\partial_{\mu} \varphi)^2 + ... \right].  
\end{equation} 
In the local potential approximation (LPA), higher derivative terms are neglected 
and the wavefunction renormalization is set equal to a
constant, i.e., $Z_k \equiv 1$. 

The solution of the RG equations sometimes requires further approximations, e.g.,
the Taylor or Fourier series of the potential $V_k(\varphi)$ in terms of the field 
variable  (with a truncation $N_{\rm cut}$)
\begin{subequations}
\begin{align}
V_{k}(\varphi) =& \; \sum_{n=1}^{N_{\rm cut}} \, \frac{g_{n}(k)}{n!} \, \varphi^{n} \,,
\\
\label{exptrig}
V_{k}(\varphi) =& \; \sum_{n=1}^{N_{\rm cut}} \, u_{n}(k) \, \cos(n \beta \varphi),
\end{align}
\end{subequations}
where the scale-dependence is encoded in the coupling constants $g_{n}(k)$
or $u_{n}(k)$.

A third systematic possibility to expand the RG flow
is provided by the so-called amplitude expansion which is frequently used.
Let us write the RG equation obtained in LPA in the following form,
\begin{subequations}
\label{lpa}
\begin{align}
k \partial_k {V}_k =& \; - \alpha_d k^d 
\int_{0}^{\infty} \dd y \,
\frac{dr}{dy} \, \frac{y^{d/2+1}}{P^2 + \omega} \,,
\\[0.133ex]
r(y) =& \; \frac{R_k( p)}{p{^2}},
\quad
y= \frac{p^2}{k^2} \,,
\quad
P^2 = [1+r(y)] \, y \,,
\quad
\omega = \frac{V''_k}{k^2} \,.
\end{align} 
\end{subequations}
The $d$-dimensional solid angle reads as $\Omega_d = 2 \pi^{d/2}/\Gamma(d/2)$,
while $\alpha_d = \Omega_d/(2(2\pi)^d)$ and $r(y)$ is 
the dimensionless regulator function. The amplitude expansion reads as
\begin{align}
k \, \partial_k {V}_k =& \;
\sum_{m=1}^\infty \frac{2m}{d} \,\, a_{2m-d} \,\, (-\omega)^{m-1} \,, 
\\
a_n =& \;
\int_{0}^{\infty} \dd y 
\left(-\frac{d}{2}\frac{r'}{(1+r)^{d/2+1}}\right) P^{-n}.
\end{align} 
When using the Litim--Pawlowski optimization \cite{opt_rg,opt_func},
the optimal choice for the parameters of the regulator functions can be 
determined in such a way to provides us the most favorable convergence of the 
amplitude expansion \cite{opt_rg,litim_o(n)}. Among the regulators,
Litim's optimized one~\cite{opt_rg} is seen to lead to the 
fastest convergence of the amplitude expansion.
Here and in the following sections of 
this article, the ``optimum'' parameters are always to be 
understood in terms of the additional approximations 
employed in the optimization process, e.g., the LPA.
The caveat is that parameters 
which are determined  as optimal within a specific, leading-order
approximation to the RG flow, are implicitly 
assumed to approximate the optimum parameters within 
different and more detailed approximation schemes.
Without this assumption, or a variation of this 
assumption, the determination 
of ``optimum'' parameters within any approximation to the RG flow
would not be meaningful. We adopt the implicit 
assumption and proceed accordingly.
The most general definition of an ``optimized RG flow''
would otherwise encompass the 
``shortest'' RG trajectory in theory space, 
where ``short'' trajectories are
quantified in terms of criteria for the gap in the flow equation.  
This sense of optimization is independent of any approximation
scheme, see Ref.~\cite{opt_func}. 
In the simple case of a derivative expansion to
lowest order, this boils down to an optimization of the 
regulator function which is the subject of the current work.

\subsection{Regulator functions}
\label{sec_regulators}
Various choices for regulator functions have already been discussed in the 
literature in a great detail. For example, one of the most frequently used 
regulator functions is the exponential one \cite{We1993}
\begin{equation}
\label{exp}
r_{\mr{exp}}(y) = \frac{a}{\exp\left(c_2 y^b \right) -1}
\end{equation} 
with $b\geq 1$. Within the LPA, a favorable choice for the parameters has been
determined as $a=1$, 
$c_2 =\ln(2)$ and $b=1.44$, based on the Litim--Pawlowski 
method~\cite{opt_rg,opt_func}. 
Another example is the power-law type regulator \cite{Mo1994}
\begin{equation}
\label{pow}
r_{\mr{pow}}(y) = \frac{a}{y^b } \,,
\end{equation} 
with $a=1$ and $b=2$ as optimal choices. Finally, one of the most 
popular regulators is the Litim choice~\cite{opt_rg}, 
\begin{equation}
\label{opt_gen}
r_{\mr{opt}}^{\mr{gen}}(y) = a \left(\frac{1}{y^b} -1\right) \Theta(1-y)
\end{equation} 
which is a continuous (but not differentiable) function with compact 
support [the Heaviside step function is denoted as $\Theta(y)$]. 
The parameters $b=1$ and $a=1$ are obtained as a 
result of the Litim--Pawlowski optimization method in LPA.

The recently introduced CSS regulator \cite{css} is defined as
\begin{equation}
\label{css_gen}
r_{\mr{css}}^{\mr{gen}}(y) = 
\frac{\exp[c \, y_0^{b}/(f-h y_0^{b})] -1}%
{\exp[c \, y^{b}/(f -h y^{b})] -1} \, 
\Theta(f-h y^b) \,.
\end{equation} 
It is infinitely differentiable 
at $f -h y^{b} =0$ in view of the essential 
singularity of the exponential function at infinity.
Let us note, that the number 
of free parameters in \eq{css_gen} can be reduced by setting $f=1$ 
without loss of generality, 
\begin{equation}
\label{css_gen_f1}
r_{\mr{css}}^{\mr{modif}}(y) = 
\frac{\exp[c y_0^{b}/(1-h y_0^{b})] -1}{\exp[c y^{b}/(1 -h y^{b})] -1}  
\Theta(1-h y^b).
\end{equation} 
Before we go on to discuss the properties of the smooth regulators, 
we should remark that that a similar smoothing problem in nuclear physics has already 
been solved by introducing the so-called Salamon-Vertse potential which 
can be related to the CSS regulator~\cite{sv}. It is easy to show that both 
forms~\eqref{css_gen} and~\eqref{css_gen_f1} of the CSS regulator have the 
property to recover all major types of regulators: Litim's optimized~\eq{opt_gen}, 
the power-law~\eq{pow} and the exponential~\eq{exp} ones,
\begin{subequations}
\begin{align}
\lim_{c\to 0, f=h=1} r_{\mr{css}}^{\mr{gen}} =  & \;
\lim_{c\to 0,h=1} r_{\mr{css}}^{\mr{modif}} = 
\frac{y_0^b \, (y^{-b}-1)}{1- y_0^b} \, \Theta(1-y), 
\\
\lim_{f\to \infty} r_{\mr{css}}^{\mr{gen}} =& \;
\lim_{h\to 0, c \to 0} r_{\mr{css}}^{\mr{modif}} = 
\frac{y_0^b}{y^b}, 
\\ 
\lim_{h\to 0,c=f} r_{\mr{css}}^{\mr{gen}}(y) =& \;
\lim_{h\to 0, c \to 1} r_{\mr{css}}^{\mr{modif}} = 
\frac{\exp[y_0^b]-1}{\exp[y^b]-1}. 
\end{align}
\end{subequations}
In order to  further reduce the number of free parameters of the CSS 
regulator, $y_0$ can be chosen so that the numerator 
of \eq{css_gen_f1} becomes a linear function of the parameter $c$
\begin{equation}
\label{css_norm1}
r_{\mr{css}}^{\mr{norm1}}(y) = 
\frac{c}{\exp[c y^{b}/(1 -h y^{b})] -1}  
\Theta(1-h y^b).
\end{equation} 
This normalization was used 
in Ref.~\cite{qeg_on_sanyi} 
in order to study the critical behaviour of 
the three-dimensional $O(N=1)$ scalar model in LPA and 
in order to investigate the
scheme-dependence of RG equations in Quantum Einstein Gravity~\cite{qeg}. 
Equation~\eq{css_norm1} has  the following limits,
\begin{subequations}
\label{css_norm1_limits}
\begin{align}
\label{css_norm1_litim}
\lim_{c\to 0,h\to 1} r_{\mr{css}}^{\mr{norm1}} = & \;
\left(\frac{1}{y^b} -1\right) \Theta(1-y), \\
\label{css_norm1_power}
\lim_{c\to 0, h \to 0} r_{\mr{css}}^{\mr{norm1}} = & \;
\frac{1}{y^b}, \\ 
\label{css_norm1_exp}
\lim_{c\to 1, h \to 0} r_{\mr{css}}^{\mr{norm1}} = & \;
\frac{1}{\exp[y^b]-1}.
\end{align}
\end{subequations}
These relations identify 
the ``Litim limit'' $c\to 0,h\to 1$,
the ``power law limit'' $c\to 0,h\to 0$,
and the ``exponential limit'' $c\to 1,h \to 0$
of the propagator.
The normalization~\eq{css_norm1} corresponds to one 
of the simplest possible choices for $y_0$,
and a CSS regulator of this kind (we refer to it as the CSS
regulator with linear norm) recovers the optimized power-law 
(with $b=2$) and optimized Litim (with $b=1$) regulators but it cannot 
reproduce the exponential regulator 
given in Eq.~\eqref{exp} with optimal parameters
[$c_2 = \ln(2)$]. 

However, 
one can choose a different normalization (let us refer to this as the 
CSS regulator with exponential norm),
\begin{align}
\label{css_norm2}
r_{\mr{css}}^{\mr{norm2}}(y) = \;
\frac{\exp[\ln(2) c]-1}{\exp\left[\frac{\ln(2) c y^{b}}{1 -h y^{b}}\right] -1}  
\Theta(1-h y^b)
= \; \frac{2^c -1}{2^{\frac{c \, y^{b}}{1 -h y^{b}}} -1} \,
\Theta(1-h y^b),
\end{align}
where the limits are
\begin{subequations}
\label{css_norm2_limits}
\begin{align}
\label{css_norm2_litim}
\lim_{c\to 0,h\to 1} r_{\mr{css}}^{\mr{norm2}} = & \;
\left(\frac{1}{y^b} -1\right) \Theta(1-y), \\
\label{css_norm2_power}
\lim_{c\to 0, h \to 0} r_{\mr{css}}^{\mr{norm2}} = & \;
\frac{1}{y^b}, \\ 
\label{css_norm2_exp}
\lim_{c \to 1, h \to 0} r_{\mr{css}}^{\mr{norm2}} = & \;
\frac{1}{\exp[\ln(2) y^b]-1}.
\end{align}
\end{subequations}
The advantage of this type of normalization is that
the form~\eq{css_norm2} reproduces all the major types of regulators 
with optimal parameters including the exponential one. This type of CSS
regulator has also been studied in
Ref.~\cite{opt_d1_d2} beyond the LPA.

Here, we consider both forms of the CSS regulator \eq{css_norm1}
and \eq{css_norm2}. The goal is to find the optimal choice of the 
parameters in LPA by using the PMS method in the framework of 
two different models in two different dimensions.

\subsection{QED${{}_2}$}
\label{sec_qed2}
One of the advantages of considering two-dimensional field theories is the
existence of bosonization transformations. Hence, gauge and fermionic 
models can be equivalently rewritten \cite{qed2,qed_qcd} in terms of scalar
fields, and in many cases, one can map the theories onto
sine-Gordon type models~\cite{sg}.
For example, we may consider 
QED$_2$ with a massive Dirac fermion which is also called 
the massive Schwinger model 
\begin{equation}
\label{qed_2}
{\cal{L}}_{\mathrm{QED_2}} = {\bar\psi}
\left(i \gamma^{\mu} \partial_{\mu} - m - e\gamma^{\mu}  A_{\mu} \right)
\psi -\frac{1}{4} F_{\mu\nu} F^{\mu\nu} \,,
\end{equation}
where $F_{\mu\nu} = \partial_{\mu}A_{\nu} - \partial_{\nu} A_{\mu}$ 
is the field-strength tensor ($\mu,\nu=0,1$). 
This model can be 
mapped onto an equivalent Bose form,
namely, the specific 
form of the massive sine-Gordon (MSG) model~\cite{qed2} whose Lagrangian 
density is written as 
\begin{equation}
\label{msg}
{\cal{L}}_{\mathrm{MSG}} = \hf (\partial_{\mu} \varphi)^2
+ \hf M^2 \varphi^2
+ u \cos (\beta \varphi) \,.
\end{equation}
Here, $\beta^2= 4\pi$, $M^2 = e^2/\pi$, 
$u = e \, m\, \exp{(\gamma_E})/(2\pi^{3/2})$,
$\gamma = 0.57721\dots$ is the Euler-Mascheroni constant,
and the vacuum angle 
parameter (a byproduct of the bosonization) 
is conveniently chosen as $\theta =\pm \pi$ for $u>0$ and $\theta =0$ 
for $u<0$ (see Ref.~\cite{qed2}). 
The MSG model has two phases~\cite{qed2}. The Ising-type 
phase transition is controlled by the dimensionless quantity $u/M^2$ related to 
the critical ratio $(m/e)_c$ of QED$_2$ which separates the confining and the 
half-asymptotic phases of the fermionic model. 

The critical ratio $0.13 < (m/e)_c < 0.33$ has been calculated by the density matrix 
RG method for the fermionic model \cite{dmrg_critical},
implying that $0.156 < [u/M^2]_c  < 0.168$.
In the framework of functional RG,
the preferred result in LPA for the critical ratio reads 
$[u/M^2]_c = 2/(4 \pi) \approx 0.15915$.
This result can be determined by analytic 
considerations based on the infrared (IR) limit  of the propagator, 
$\lim_{k\to 0} (k^2 + V''_k(\varphi)) = 0$,
where $V_k(\varphi)$ is the blocked scaling 
potential which contains the mass term and all the higher harmonics 
generated by RG transformations \cite{qed2}. However, if one considers 
a single Fourier-mode approximation 
[where $V_k(\varphi)$ contains the mass term and 
only a single cosine], then
the analytic result based on the IR behavior of the 
propagator~\cite{IR,qed2} gives 
\begin{eqnarray}
\label{ratio}
\chi_c = 
\left[\frac{u}{M^2}\right]_c = \frac{1}{(4 \pi)} \approx 0.07957.  
\end{eqnarray}
In this case,
the optimized regulator~\eq{opt_gen} with $b=1$ and $c = 0.01$ 
leads to a ratio $[u/M^2]_c = 0.07964$ closer to the analytic one
\cite{qed2} but other regulators such as the power-law type regulator with
$b=1$ run into a singularity and stop at some finite momentum scale,
rendering 
the determination of the critical ratio~\cite{qed2} impossible. Therefore,
the use of the single Fourier mode approximation provides us 
with a tool to
consider the convergence properties of the RG equations and to optimize the
regulator functions. Here, we use the PMS method in order to determine the
optimized parameters of the CSS regulator with various
normalizations [see Eqs.~\eq{css_norm1} and \eq{css_norm2}]
and compare the
ratio given by the CSS regulator with these optimized parameters to the
analytic result~\eqref{ratio} which represents the optimum
available ratio in the case
of a single-Fourier-mode approximation.

%
%
\begin{figure}[thb] 
\begin{center} 
\includegraphics[width=0.42\linewidth]{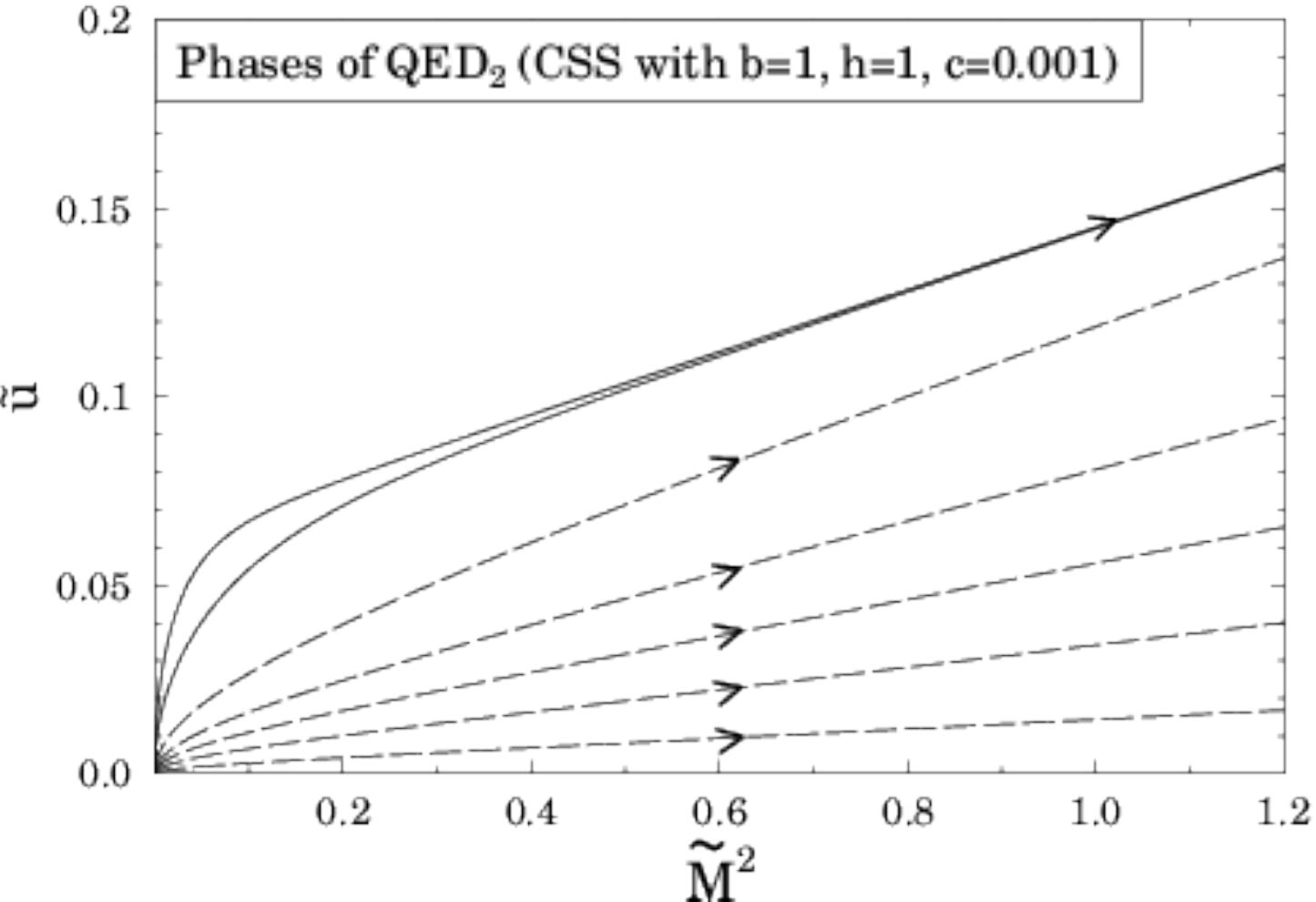}
\begin{minipage}{0.8\textwidth}
\caption{\label{fig1}
Phase diagram of the MSG model for $\beta^2 =4\pi$ is obtained by 
the usage of the CSS regulator \eq{css_norm1} with $b=1$, $h=1$ and $c=0.001$. 
RG trajectories of the broken symmetric phase (full lines) merge into a single one 
and the critical ratio of the model is determined by its slope in the IR limit. The arrows 
indicate the direction of the flow.} 
\end{minipage}
\end{center}
\end{figure}

%
%
\begin{figure}[thb] 
\begin{center} 
\includegraphics[width=0.42\linewidth]{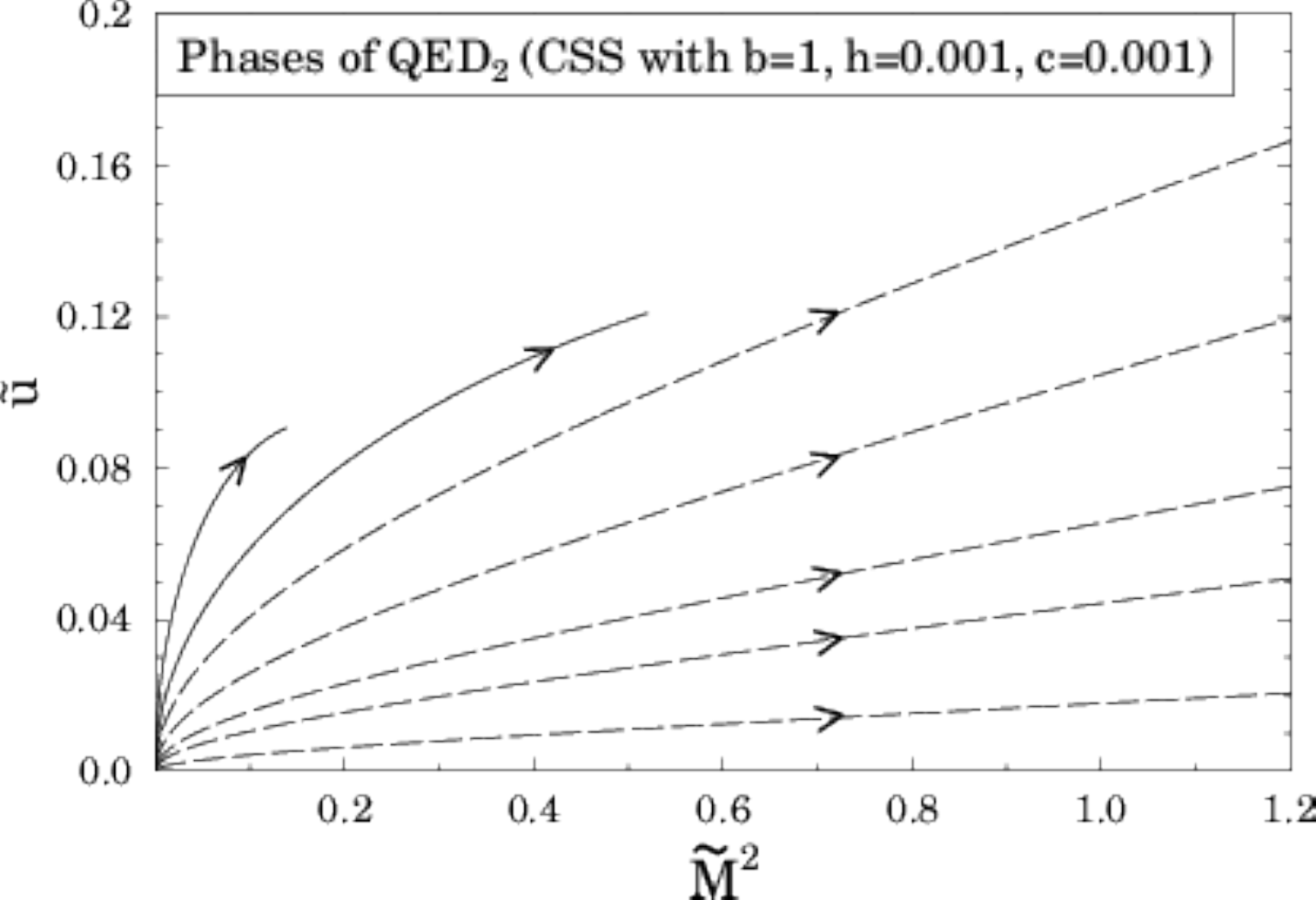}
\begin{minipage}{0.8\textwidth}
\caption{\label{fig2}
Phase diagram of the MSG model for $\beta^2 =4\pi$ is obtained by 
the usage of the CSS regulator \eq{css_norm1} with $b=1$, $h=0.001$ and 
$c=0.001$, approximating the power-law limit. 
RG trajectories of the broken symmetric phase (full lines) 
stop at some finite momentum scale $k_{\rm f}$ before they  
merge into a single one,
and the determination of the critical ratio is not possible.} 
\end{minipage}
\end{center}
\end{figure}

%
%
\begin{figure}[htb] 
\begin{center} 
\includegraphics[width=0.9\linewidth]{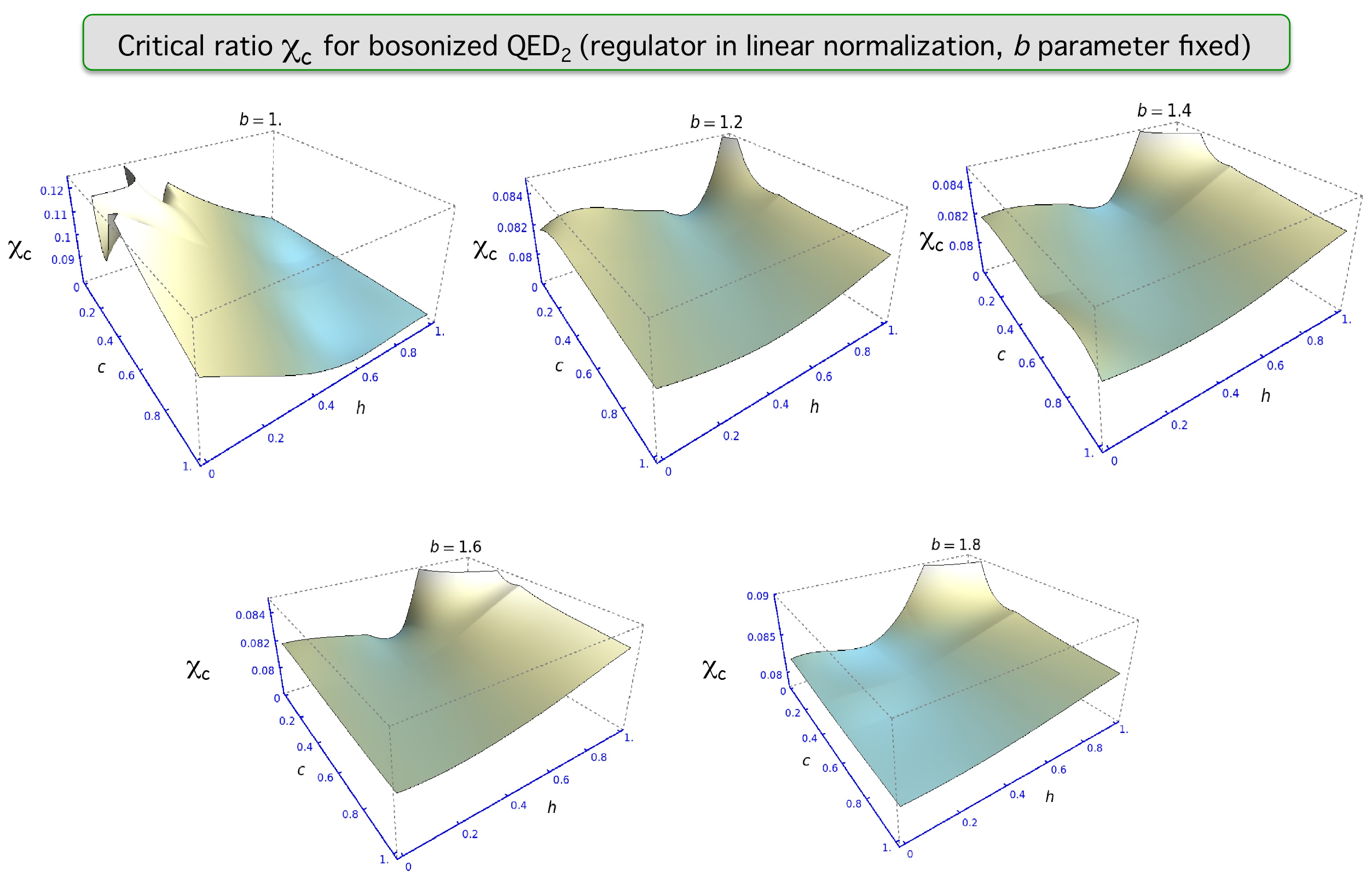}
\caption{\label{fig3}
The critical ratio $\chi_c$, defined 
according to Eq.~\eq{ratio}, 
of bosonized QED$_2$, obtained with the
CSS regulator in the ``linear'' normalization~\eq{css_norm1} for various parameters. 
Lower ratios indicate better regulators. The optimum result is obtained
for $b=1$, $c=0.001$ and $h=1$ [``back corner'' of Fig.~(a)]
which is numerically close 
to the Litim limit of the CSS, according to Eq.~\eqref{css_norm1_litim}.
The critical ratio $\chi_c$ is on the ordinate axis.} 
\end{center}
\end{figure}

\section{Results obtained by the RG study of QED$_2$}
\label{sec_results_qed2}
\subsection{Orientation}
We consider the RG study of the bosonized QED$_2$ 
described by the Lagrangian~\eqref{msg}. We use 
two types of approximations, 
{\em (i)} RG equations are obtained within the LPA
[see Eq.~\eq{lpa}], 
{\em (ii)} a single-mode Fourier approximation. 
In the latter case, the dimensionless 
blocked potential ($\tilde V_k = k^{-2} V_k$) with a single Fourier mode 
is used in the form
\begin{eqnarray}
\label{single_msg}
\tilde V_k = \tfrac12 \, {\tilde M}_k^2 \, \varphi^2 + 
{\tilde u}_k \; \cos (\beta \varphi)\,,
\end{eqnarray}
with $\beta^2 = 4\pi$,
and the tilde superscript denotes the dimensionless 
couplings, ${\tilde M}^2 = k^{-2} M^2$ and ${\tilde u}_k = k^{-2} u_k$. This
is inserted into the dimensionless form of Eq. \eq{lpa} in $d=2$ which
reads as
\begin{eqnarray}
\label{dimless_rg}
(2+k\partial_k) {\tilde V}_k(\varphi) = - 
\int_0^\infty \frac{\dd y}{4 \pi} \,
\frac{dr}{dy}
\frac{y^2}{(1+r)y + {\tilde V''}_k(\varphi)}.
\end{eqnarray}
We emphasize that no further approximations such as the 
amplitude expansion discussed in Sec.~\ref{sec_approx} are used, here.

The phase structure of the single-frequency MSG model \eq{msg},
as given in~Fig.~\ref{fig1}, 
can be obtained from the functional RG equation~\eq{dimless_rg}
using the CSS regulator in 
``normalization 1'' \eq{css_norm1} with $b=1$, $h=1$ and $c=0.001 \approx 0$.
If the CSS regulator is normalized according to Eq.~\eq{css_norm2}
(``normalization 2''), then
almost identical results are obtained because both 
normalizations~\eq{css_norm1} and~\eq{css_norm2},
tend to Litim's optimized form in the 
limit $c\to 0$ (this was also demonstrated in Ref.~\cite{css}).
All numerical calculations reported here 
have been obtained via a combination of {\tt C++} codes
and computer algebra software~\cite{Wo1999}.

The RG trajectories represented by full lines in Fig.~\ref{fig1} correspond
to the broken phase of the single-frequency MSG model \eq{msg} 
where the reflection symmetry ($Z_2$) is broken spontaneously. These
RG trajectories merge into a single one in the IR limit and its slope defines 
the critical ratio. Thus, in the broken phase 
$\tilde u_k$ is a linear function of $\tilde M_k^2$ in the IR limit,
\begin{eqnarray}
\label{linear}
\tilde u_{k} = a \,\, \tilde M_{k}^2 + b.
\end{eqnarray}
Both the ratio of the dimensionful coupling to the mass term,
as well as the dimensionless equivalent tend to the constant
$[{\tilde u}_{k\to 0}/{\tilde M}_{k\to 0}^2] = [u_{k\to 0}/M_{k\to 0}^2] = a$
(since $\tilde u_{k}$ and $ \tilde M_{k}^2$ are increasing in the IR limit,   
their ratio tends to $a$ and the constant $b$ term can be neglected),
and the slope is independent of the initial conditions 
(in the symmetric phase, the linear 
functional form \eq{linear} holds but the slope depends on the initial values). 

%
%
\begin{figure}[thb] 
\begin{center} 
\includegraphics[width=0.42\linewidth]{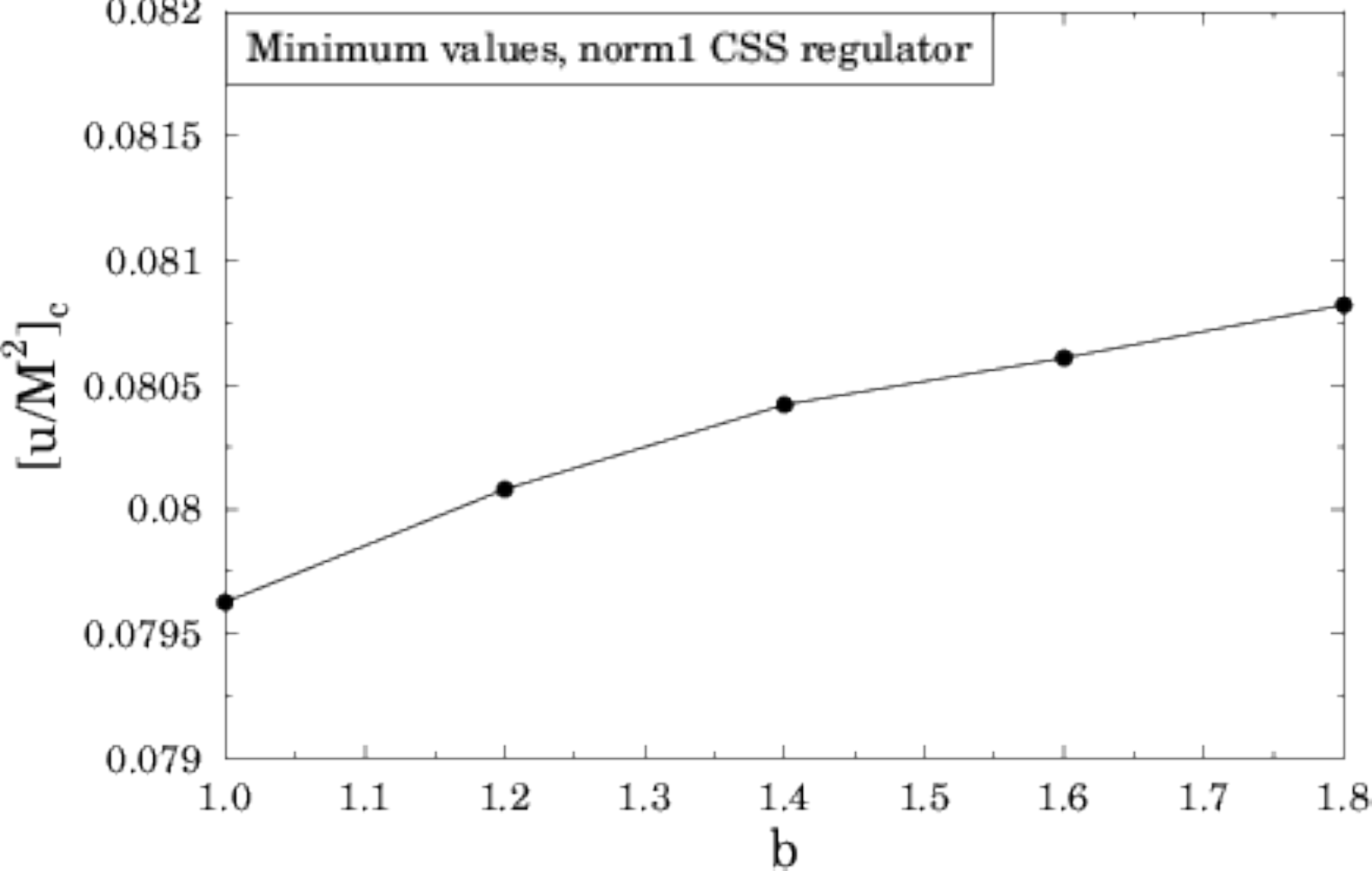}
\begin{minipage}{0.8\textwidth}
\caption{\label{fig4}
Minimum values (with respect to $c$ and $h$ for a
fixed $b$) of the critical ratio obtained by the CSS regulator with linear norm 
are shown as a function of the parameter $b$. It demonstrates that 
the limit $b\to 1$, $c\to 0$ and $h=1$ gives the numerically
most favorable results.}
\end{minipage}
\end{center}
\end{figure}

There is strong numerical evidence, obtained in this work,
for a scheme-independent slope $a = 1/(4\pi)$,
within the single Fourier-mode approximation.
However, the convergence properties of the RG equation 
depend on the regulator chosen. The RG evolution stops at some finite scale 
$k_{\rm f}\neq 0$ and the ratio $[u_{k\to k_{\rm f}}/M_{k\to k_{\rm f}}^2]$ becomes 
scheme-dependent, where the optimization can be performed. 
This strategy is used 
to select the optimized regulator according to its convergence properties. As an
example for a special situation,
one has to mention the power-law limit ($h\to 0$, 
$c\to 0$) of the CSS regulators
given in Eqs.~\eq{css_norm1_power} and \eq{css_norm2_power}.
The power-law regulator has poor convergence properties
for $b=1$, and thus the RG trajectories of the broken phase cannot merge into a single
one and the determination of the critical ratio is not possible at all. This is shown
in~Fig.~\ref{fig2} for the CSS regulator \eq{css_norm1_power} with $b=1$, 
$h=0.001 \approx 0$ and 
$c=0.001 \approx 0$, approximating the power-law limit. 
The same conclusion can be obtained
on the basis of Eq.~\eq{css_norm2_power}.

\subsection{QED${{}_2}$ and CSS regulator with linear norm}
We first attempt to find the optimal set of parameters $b$, $h$, $c$ for the CSS 
regulator  with the 
linear norm~\eq{css_norm1}. In~Fig.~\ref{fig3}, the critical ratio 
$[u/M^2]_c = \tilde u_{k\to k_f} /\tilde M_{k\to k_f}^2 = u_{k\to k_f} /M_{k\to k_f}^2$ 
obtained by the CSS with linear norm is shown as a 
function of $b$, $h$ and $c$.  
The lower the ratio the better the regulator is. 
We obtain results closest to the 
analytical value \eq{ratio} for 
the parameters $b=1$, $c=0.001$ and $h=1$,
which are very close to the Litim limit of the CSS. 
Let us apply the PMS strategy in order to determine
the  optimized set of parameters of the CSS regulator which is chosen by finding 
the extremum (in this particular case the minimum) of a given physical parameter 
(for QED$_2$ the critical ratio $[u/M^2]_c$)
obtained by the RG method. Thus, near
the optimized parameters, the results obtained by the RG study of the model has 
a minimal sensitivity on the change of parameters. For example, one finds a 
plateau in Fig.~\ref{fig3} for every subgraph (where the parameter $b$ is fixed) which 
defines the optimized parameters. 
For $b=1$,
the optimized value as a function of the 
two remaining arguments is the Litim limit $c\to 0$, $h=1$.
For $b\neq 1$, the optimized parameters are found in the limit
$c\to 0$, but at $h\neq 1$, still leading to less 
effective global optimization of the 
critical ratio.

This can be seen in~Fig.~\ref{fig4}, where we plot the minimum values for the
critical ratio obtained at every subgraph of~Fig.~\ref{fig3} in terms of the
remaining parameters $c$ and $h$, for ``fixed'' value of $b$.  The dependence
of each minimum, found as a function of $c$ and $h$, can thus be read off as a
function of $b$, demonstrating that the limit $b\to 1$, $c\to 0$ and $h=1$
leads to the globally most effective optimization.

%
%
\begin{figure}[htb]
\begin{center}
\includegraphics[width=0.9\linewidth]{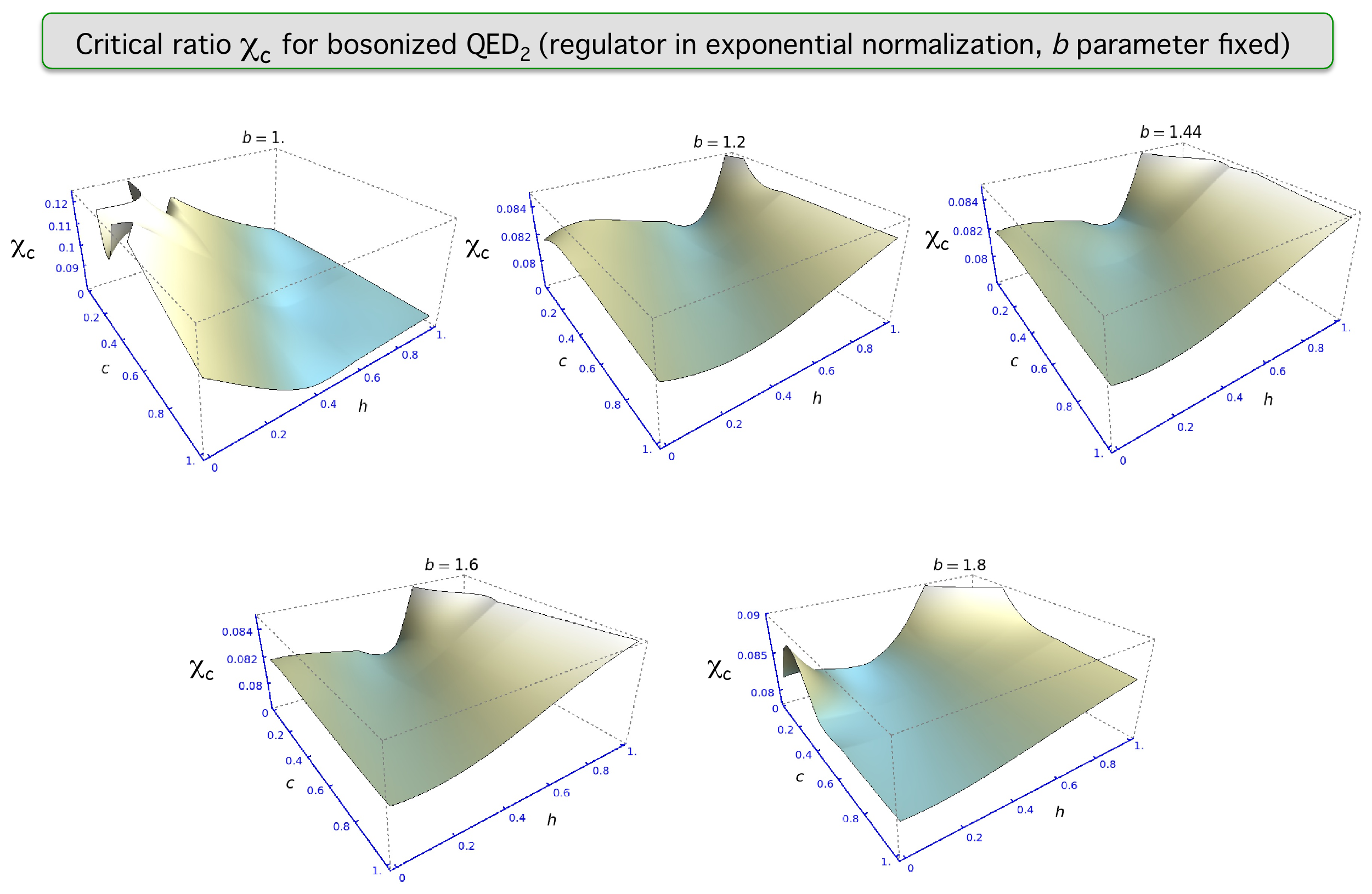}
\caption{\label{fig5}
We plot the critical ratio $\chi_c$ of  bosonized QED$_2$, obtained
by various parameters of the CSS regulator with 
exponential norm~\eq{css_norm2}. The critical ratio $\chi_c$ is on the ordinate axis.
Lower critical ratios indicate better regulators, with the optimum results 
being obtained for $b=1$, $c=0.001$ and $h=1$ (Litim
limit of the CSS). The geometry of the plots is similar to those
recorded in~Fig.~\ref{fig3} but numerical 
differences are observed. For larger 
values in the range $b \approx 1.8$, an additional local maximum 
develops as a function of $c$, along the ($h=0$)-manifold.}
\end{center}
\end{figure}

%
%
\begin{figure}[t!]
\begin{center}
\includegraphics[width=0.42\linewidth]{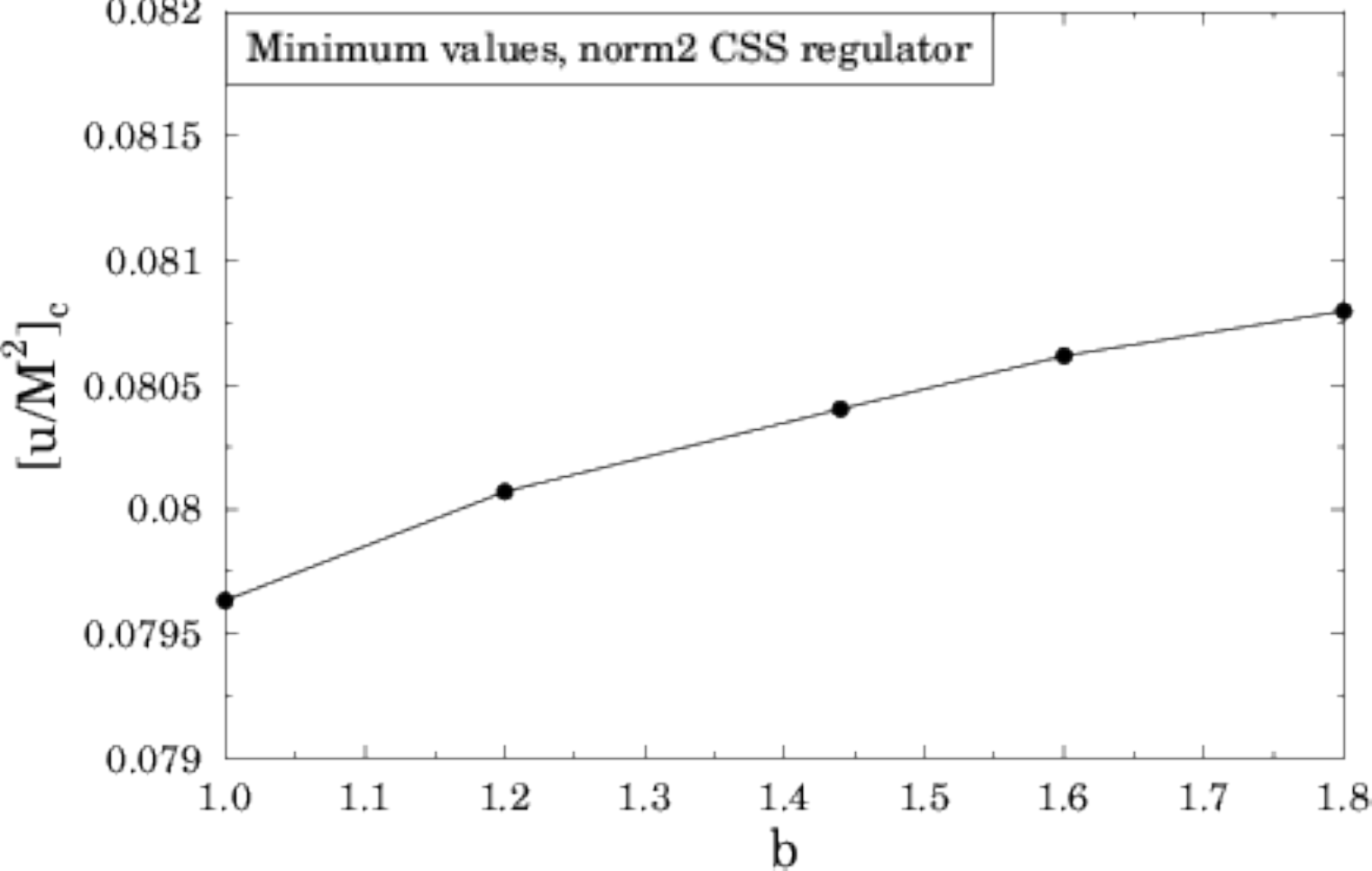}
\begin{minipage}{0.8\textwidth}
\caption{\label{fig6}
Minimum values (with respect to $c$ and $h$ for 
fixed $b$) of the critical ratio obtained by the CSS regulator with exponential
norm are plotted for various values of the parameter $b$.
Again, the most favorable results
are obtained in the limit $b\to 1$, $c\to 0$ and $h \to 1$.
Notably, the optimum critical ratio $[u/M^2]_c$ obtained 
for $b\to 1$ is lower than the ``optimum'' value 
(with respect to a variation of $c$ and $h$) 
obtained for $b = 1.44$.}
\end{minipage}
\end{center}
\end{figure}

\subsection{QED${}_2$ and CSS regulator with exponential norm}
Let us repeat the calculation done in the previous subsection with the CSS,
this time employing the exponential norm~\eq{css_norm2} instead of the linear
normalization~\eq{css_norm1}.  The goal is the same, i.e., to find the optimal
set of parameters $b$, $h$, $c$. In~Fig.~\ref{fig5}, the critical ratio
obtained by the CSS with exponential norm is shown as a function of $b$, $h$
and $c$.  In analogy to the previous analysis for linear norm, the optimum
result, i.e., the one closest to the analytic formula \eq{ratio}, is obtained
for $b=1$, $c=0.001$ and $h=1$.  This, again, is the Litim limit of the CSS
regulator.  In~Fig.~\ref{fig6}, we plot the minimum values for the critical
ratio obtained at every subgraph of Fig.~\ref{fig5}, i.e., keeping $b$ fixed,
and thus the dependence of each minimum on the parameter $b$ can be read off,
confirming once more the nature of the optimized propagator.  Again, this
demonstrates that Litim's limit ($b\to 1$, $c\to 0$ and $h=1$) leads to the
optimum results.  The PMS strategy gives us the same optimal choice for the
parameters of the CSS regulator, the limit $b\to 1$, $c\to 0$ and $h=1$, 
independent of the
normalization of the CSS regulator.  In order to gain more accurate information
on the optimal choice of the parameters for the CSS regulators \eq{css_norm1}
and \eq{css_norm2}, we now consider the $O(N=1)$ symmetric
scalar field theory in $d=3$ dimensions (which constitues 
a ``textbook example'').

\section{Three-dimensional $\bm{O(1)}$ model}
\label{sec_phi4}
\subsection{Orientation and previous studies}
A further textbook example for the optimization of the 
parameters entering the propagator is given by the $O(1)$ model in 
$d=3$ dimensions around the Wilson-Fisher (WF) fixed point and the 
corresponding critical exponent~$\nu$ (which describes the
critical behaviour of the correlation length). Similar to 
the study of the bosonized QED$_2$ with a single Fourier mode, the
$O(1)$ model is investigated 
here using a drastic truncation (as drastic as possible),
i.e., the dimensionless potential is defined as 
\begin{equation}
\label{ansatz}
\tilde V = \frac{1}{2} \, \tilde g_1 \; \tilde \varphi^{2} 
+ \frac{1}{4!} \, \tilde g_2 \; \tilde \varphi^{4},
\end{equation}
with the dimensionless couplings $\tilde g_1$ and $\tilde g_2$. 
With this truncation, the RG flow equations obtained for the couplings 
read as~\cite{qeg_on_sanyi},
\begin{eqnarray}
\dot{\tilde g}_1 &=& -2 \tilde g_1 + \tilde g_2 \; 
\bar \Phi^2_{3/2}(\tilde g_1), \nn
\dot{\tilde g}_2 &=& -\tilde g_2 + 6 \, \tilde g_2^2 \;
\bar \Phi^3_{3/2}(\tilde g_1),
\end{eqnarray}
where the threshold function is introduced according to
\begin{equation}
\bar \Phi^p_n(\omega) = \frac{1}{(4\pi)^n \, \Gamma(n)}
\int_0^\infty \dd y \frac{y^{n+1} \, r'}{(y(1+r)+\omega)^p} \,.
\end{equation}
The flow equations obviously are regulator-dependent, and the exponent $\nu$
calculated near the WF fixed point is regulator-dependent too.  The threshold
function has been calculated for all major types of regulators in the
literature, see e.g. \cite{scheme,opt_rg,opt_func,litim_o(n)}.  Among these
regulators, the Litim regulator is known to give us the ``best'' critical
exponent (closest to the ``exact'' value). For two couplings, the result for
$\nu$ obtained using the Litim form of the regulator reads as $\nu \approx
0.54277$ \cite{qeg_on_sanyi}.

The CSS regulator with the linear norm~\eq{css_norm1} has been studied in
Ref.~\cite{qeg_on_sanyi} using the RG evolution for 2 and 4 couplings, where it
has been confirmed that the optimized parameters of the CSS with linear norm
are those of the Litim limit., supporting the observations made in
Sec.~\ref{sec_results_qed2}.  However, the CSS with exponential norm has not
yet been studied in the framework of the $O(1)$ model, so we perform a detailed
analysis of the CSS regulator in the normalization~\eq{css_norm2} looking for
the optimal set of parameters. 

%
%
\begin{figure}[t!]
\begin{center}
\includegraphics[width=0.9\linewidth]{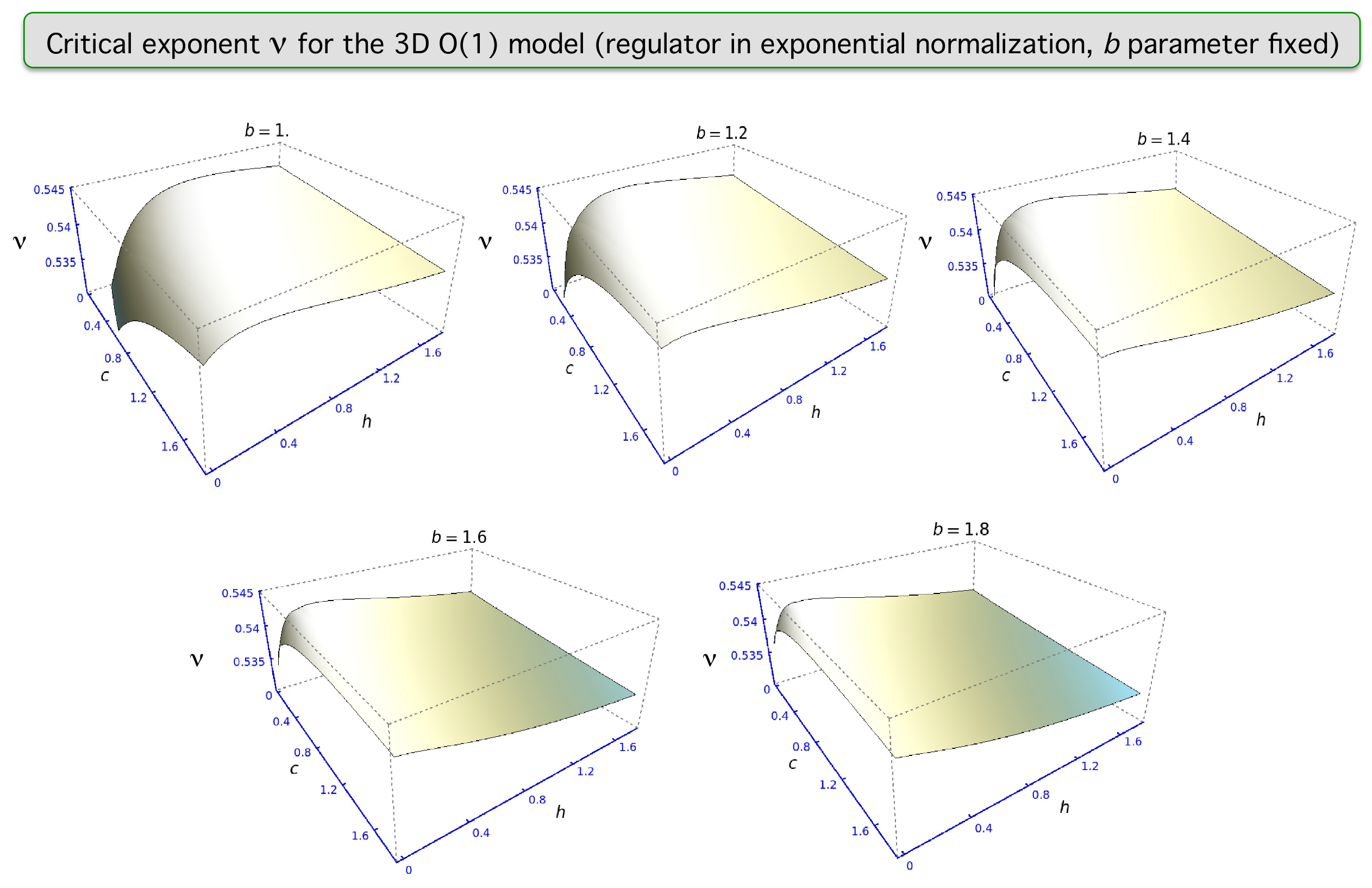}
\caption{\label{fig7}
The critical exponent $\nu$ of the three-dimensional $O(1)$ model with
two couplings, as given by the potential of
Eq.~\eq{ansatz}, is obtained by various parameters of the CSS regulator
with exponential norm~\eq{css_norm2}. Higher exponents indicate better
regulators. The critical exponent $\nu$ is plotted 
on the ordinate axis.}
\end{center}
\end{figure}
%

%
%
\begin{figure}[t!]
\begin{center}
\includegraphics[width=0.42\linewidth]{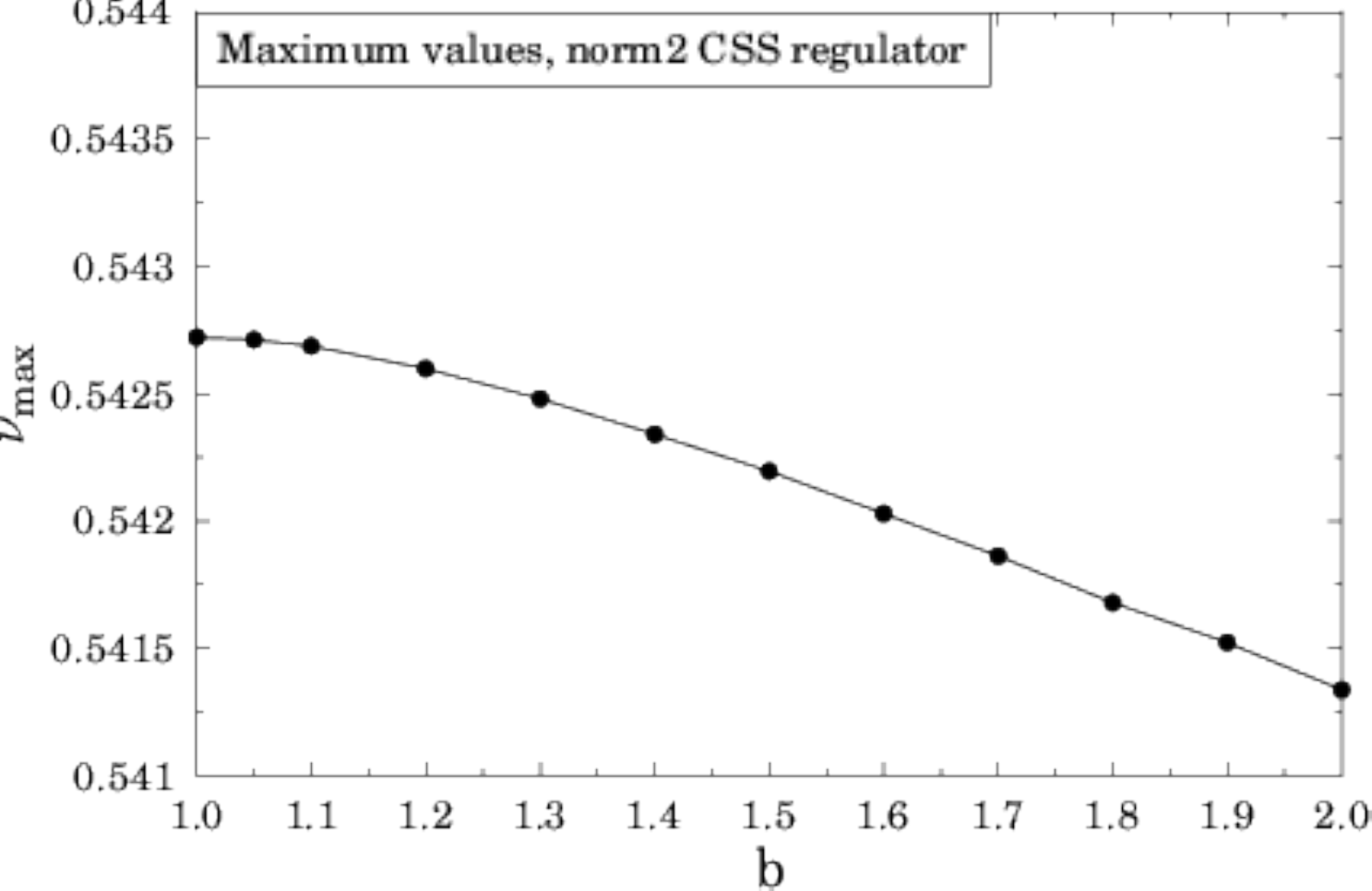}
\begin{minipage}{0.8\textwidth}
\caption{\label{fig8}
The maximum values of the critical exponent $\nu$ obtained by the CSS
regulator with exponential norm~\eqref{css_norm2} for various values of the
parameter $b$ is shown. Using the PMS method, one finds that the Litim limit
$b\to 1$ leads to the optimum result.}
\end{minipage}
\end{center}
\end{figure}

%
%
\begin{figure}[t!]
\begin{center}
\includegraphics[width=0.42\linewidth]{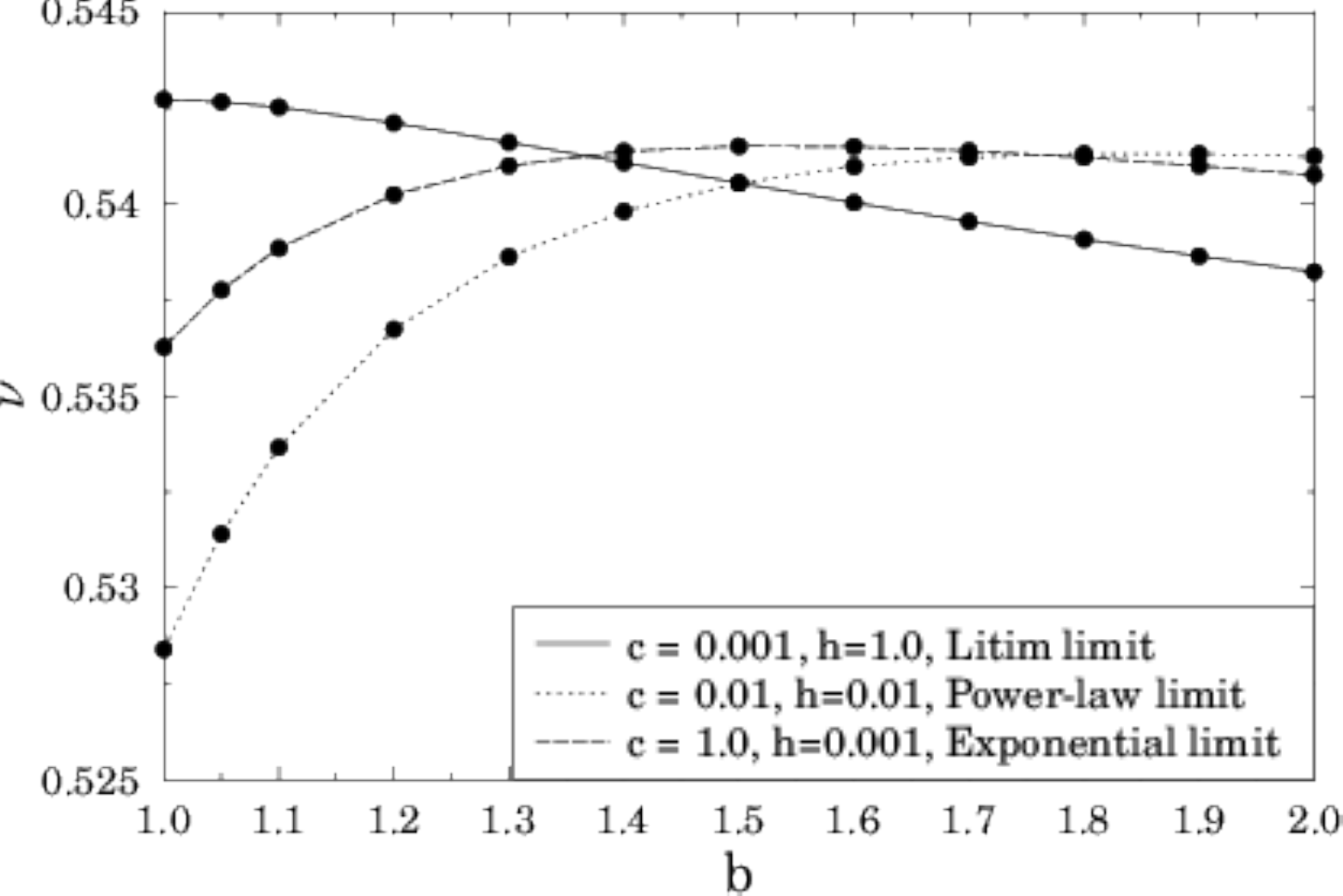}
\begin{minipage}{0.8\textwidth}
\caption{\label{fig9}
The $b$-dependence of the critical exponent $\nu$ obtained by
the CSS regulator with exponential norm~\eqref{css_norm2}  in various limits is shown.}
\end{minipage}
\end{center}
\end{figure}
%

\subsection{Three-dimensional $\bm{O(1)}$ model and CSS regulator with exponential norm}
In this subsection, we perform the RG study of the 
$O(1)$ model~\eq{ansatz} by
the CSS regulator with exponential norm \eq{css_norm2}. Our goal here
is the same as it was in the case of QED$_2$, namely, we are looking for the 
optimal set of parameters ($b$, $h$, $c$). Here, the critical exponent $\nu$ of 
the $O(1)$ model (with 2 couplings) is determined and one finds the optimized 
parameters at which $\nu$ is maximal. In~Fig.~\ref{fig7},
we plot the critical exponent $\nu$ obtained by 
the CSS regulator with exponential norm as a function of $b$, $h$ 
and $c$.  

We apply the PMS strategy.  Near the optimized set of parameters,
the results obtained for $\nu$ by the RG study of the model have
minimal sensitivity on small perturbations. Around these optimized parameters,
one finds plateaus in Fig.~\ref{fig7} for every subgraph. If we identify $\nu_{max}$
as the maximum of $\nu$ with respect to $c$ and $h$ of every subgraph of
Fig.~\ref{fig7}, and we plot it against the parameter $b$, see Fig.~\ref{fig8}, one can
read off the absolute maxima using the PMS principle. 
According to the PMS method, the optimum result is obtained for $b=1$, 
$c=0.001$ and $h=1$ which is the (numerically 
close to) Litim limit of the CSS regulator. We recall that
both normalizations~\eq{css_norm1} and~\eq{css_norm2}
converge to the Litim limit for $c \to 0$, and $h=1$.

Another perspective on the problem is obtained when we consider the CSS
regulator for fixed values of the parameters $c$ and $h$, and investigate the
critical exponent $\nu$ as a function of $b$.  The results obtained by the CSS
regulator regulator with exponential norm \eq{css_norm2} in the Litim,
power-law and exponential limits are plotted in Fig.~\ref{fig9}. In the Litim
limit of the CSS, the optimum results obtained by the PMS strategy are obtained
for $b=1$ (which is the generally accepted value in the literature). In the
power-law limit of the CSS, one finds $b\approx 2$.  Based on the
field-amplitude expansion, it is known that the optimum choice for the
power-law regulator is $b=2$.  In the exponential limit of the
CSS~\eq{css_norm2}, the most favorable results were obtained for $b\approx 1.5$
(consistent with a visual inspection of~Fig.~\ref{fig9}), while it is known
that, based on the field-amplitude expansion, the optimum choice for the
exponential regulator is $b=1.44$. These latter observations are 
important in order to ensure the consistency of our calculations
with known results, for specific, fixed values of the regulator parameters.

\section{Summary} \label{sec_sum}
In this paper, we have discussed the optimization of the regulator-dependence
of functional RG equations in LPA, for the bosonized QED$_2$ and the
three-dimensional $O(1)$ scalar theory. The optimization has been done in the
framework of the PMS method.  It has been known
\cite{litim_o(n),opt_rg,opt_func} that Litim's regulator leads to the fastest
convergence of the field amplitude expansion. This regulator is also known as
the optimum  regulator in LPA, which yields critical exponents of the $O(N)$
model very close to the ``exact'' ones.  The recently introduced CSS regulator
\cite{css}, which has a very general functional form [see Eq.~\eq{css_gen_f1}],
has not yet been subjected to a thorough optimization analysis.  The CSS
regulator reduces to all major type of regulator functions in appropriate
limits [see Eqs.~\eqref{css_norm1_limits} and~\eq{css_norm2_limits}]. Due to
its versatile functional form, the CSS regulator represents an excellent
playground for the PMS method allows to systematically investigate the
optimization of the regulator within a rather wide class of regulator
functions. The CSS regulator has been employed here with two different
normalizations~\eq{css_norm1} and~\eq{css_norm2} (``linear'' and
``exponential'' norm).  In order to numerically study the convergence
properties of RG equations with various regulators, we have used a truncation
of the field expansion for both models, i.e., a single Fourier mode
approximation for the bosonized  QED$_2$ and a single polynomial approximation
for the $O(1)$ model in three dimensions (i.e. the inclusion of the $\varphi^4$
term in the blocked action but no higher polynomials). Our results can be
summarized as follows.

{\em (i)} Litim's form of the CSS regulator has been found to be the optimum
choice in both cases.  In particular, we have been unable to find a set of
parameters of the general CSS regulator, which would otherwise lead to
numerically more favorable results as compared to those obtained with Litim's
form.  This statements holds for both QED$_2$ as well as the three-dimensional
$O(1)$ model. 

{\em (ii)} The PMS method can be applied to the RG analysis of the CSS very
effectively.  In particular, it allows us to search for the numerically optimal
set of parameters within a three-dimensional space of parameters $b$, $h$ and
$c$ describing the general form of the CSS regulator~\eq{css_gen_f1}.  We
confirm that Litim's form of the regulator leads to the numerically most
favorable results within the LPA.  Such a general investigation based on the
PMS method has not been recorded in the literature to the best of our
knowledge.  Regarding possible future investigations, we would like to expand
that in principle, the combination of the CSS regulator and the PMS was found
to be a very powerful method which can be applied in any dimension, for any
type of models, in any order of the gradient expansion. 

{\em (iii)} Various known results were recovered by using the CSS regulator in
specific limits.  For example, in the exponential limit of the CSS, for the
three-dimensional $O(1)$-model, the optimum results for the critical exponent
$\nu$ were obtained for $b\approx 1.5$, which compares well with the
field-amplitude expansion, where the known optimized value for the exponential
regulator is $b=1.44$. 

Finally, let us note that the CSS regulator is infinitely differentiable for
any nonzero values of $c$ and $h$, while in the exact limit $c \to 0$ advocated
by Litim, a kink develops at $y=1$.  However, for any manifestly nonzero value
of $c$, say, $c = 0.001$, it can be applied to higher order terms of the
gradient expansion in any order, unlike Litim's original regulator which is
non-differentiable. This would indicate that the form of an  ``optimized CSS
regulator'' with a small, but nonzero value of $c$, remains both compatible
with the gradient expansion while reproducing the optimum values for the
critical ratio $\chi_c$ and the critical exponent $\nu$ in appropriate limits.
A nonvanishing, small but finite value of the $c$ parameter regularizes the
kink of the regulator functions (at the end point of the compact 
support of the regulator). The essential singularity of the 
exponential function at infinity is crucial in the regularization 
process. Let us study the regulator~\eqref{css_norm1} for $b=1$,
\begin{equation}
r_{\mr{css}}^{\mr{norm1}}(y) = 
\frac{c}{\exp[c \, y/(1 -h y)] - 1} \,
\Theta(1-h y) \,,
\end{equation}
with a special emphasis on the limit $c \to 0$. 
We first observe that, globally, for $c \to 0$, the
regulator retains a finite limit, because the small factor $c$ in the numerator
cancels against the same factor from the denominator.
One might ask if divergences could occur for small values of 
$c$ when higher-order derivative terms are considered.
Preliminary results, based on example calculations,
indicate that this is not the case.
Typical integrals which occur in calculations beyond the LPA,
for the sine-Gordon models in one and two dimensions, 
have been given in Eqs.~(8) and (9) of Ref.~\cite{opt_d1_d2}.
The distance $D$ between the non-trivial IR fixed point and
the saddle point of the RG flow, 
for the mentioned sine--Gordon models,
has been given in Eq.~(10) of Ref.~\cite{opt_d1_d2}.
Preliminary numerical numerical suggests
that no additional divergences occur for small values of $c$,
even if derivatives of the regulator function 
are taken according to Eqs.~(8) and (9) of Ref.~\cite{opt_d1_d2}.
This is consistent with the nonperturbative 
character of the CSS regulator near the end point of the 
compact support interval, due to the essential singularity 
of the exponential function at infinity.
(Further details will be discussed elsewhere.)
Thus, there is hope that in future studies on related problems,
the CSS regulator, with parameters close to the Litim limit,
might be very useful in advances beyond the LPA, and may
recommend a regularized, and numerically optimized form of the regulator.
Identifying the optimum form of the regulator function is
important in order to improve the general understanding  of the scheme
dependence of the functional renormalization group.

\section*{Acknowledgements}

This work was partially supported by the European Union and the European 
Social Fund through project Supercomputer, the national virtual lab 
(grant no.: TAMOP-4.2.2.C-11/1/KONV-2012-0010).
U.D.J.~acknowledges support from the National Science Foundation
(Grant PHY--1068547) and from the 
National Institute of Standards and Technology 
(precision measurement grant).\\[3ex]

\end{document}